\documentclass[12pt]{article}
\usepackage[russian,english]{babel}
\usepackage{amsthm, amsmath, amssymb, amsbsy, amsfonts, latexsym, euscript}

\newcommand{\beqn}{\begin{eqnarray}}
\newcommand{\eeqn}{\end{eqnarray}}
\newcommand{\be}{\begin{equation}}
\newcommand{\ee}{\end{equation}}
\newcommand{\ba}{\begin{array}}
\newcommand{\ea}{\end{array}}
\newcommand{\R}{{\rm\bf R}}

\newcommand{\re}{\ref}
\newcommand{\ci}{\cite}
\newcommand{\la}{\label}
\newcommand{\bfr}{\begin{flushright}}
\newcommand{\efr}{\end{flushright}}
\newcommand{\bfl}{\begin{flushleft}}
\newcommand{\efl}{\end{flushleft}}
\newcommand{\fr}{\frac}

\newcommand{\st}{\stackrel}

\newcommand{\ds}{\displaystyle}
\newcommand{\ve}{\varepsilon}

\newcommand{\na}{\nabla}
\newcommand{\lam}{\lambda}

\newcommand{\br}{|\kern-.25em|\kern-.25em|}
\newcommand{\brr}{{|\kern-.15em|\kern-.15em|\kern-.15em}\,}
\newcommand{\ddd}{\st{.\kern-.07em.\kern-.07em.}}
\def\N{{\rm I\kern-.1567em N}}                              
\def\R{{\rm I\kern-.1567em R}}                              
\def\C{{\rm C\kern-4.7pt                                    
\vrule height 7.7pt width 0.4pt depth -0.5pt \phantom {.}}}
\def\Z  {{\sf Z\kern-4.5pt Z}}                                

\begin{document}
\renewcommand{\theequation}{\thesection.\arabic{equation}}
\newtheorem{theorem}{Theorem}[section]
\renewcommand{\thetheorem}{\arabic{section}.\arabic{theorem}}
\newtheorem{definition}[theorem]{Definition}
\newtheorem{deflem}[theorem]{Definition and Lemma}
\newtheorem{lemma}[theorem]{Lemma}
\newtheorem{example}[theorem]{Example}
\newtheorem{remark}[theorem]{Remark}
\newtheorem{remarks}[theorem]{Remarks}
\newtheorem{cor}[theorem]{Corollary}
\newtheorem{pro}[theorem]{Proposition}


\begin{center}
{\Large\bf On Global Attraction for a Particle\medskip\\
Coupled to a Scalar Field}\\
\vspace{1cm}
{\large Valeriy Imaykin}

\bigskip

State Budgetary General Education Institution of the City of Moscow\\ ``School No. 179'', Moscow, Russian Federation

\medskip

email: ivm61@mail.ru

\end{center}


\begin{abstract}
We study the question of global attraction, in the energy norm, for finite
energy solutions to classical particle coupled to a scalar wave field. The attraction could take place to either the set of the stationary solutions in the case of a confining potential or to the soliton manifold in the case of zero potential.

By energy conservation argument we establish that in both cases there is no global attraction.
\end{abstract}


\setcounter{equation}{0}

\section{Introduction}

Consider a single charge coupled to a scalar wave  field and subject to
an external potential in 3-dimensional space.
If $q(t)\in\R^3$ denotes the position of charge at time $t$, then
the coupled equations read
 \beqn\la{1}
\ba{ll}
\dot{\psi}(x,t)=\pi (x,t), & \dot{\pi}(x,t)=\Delta \psi (x,t)-\rho (x-q(t)),
\\
~ &  \\
\dot{q}(t)=p(t)/(1+p^2(t))^{1/2}, & \dot{p}(t)=-\na V(q(t))+
\displaystyle\int dx\,\psi
(x,t)\,\nabla \rho (x-q(t)).
\ea
\eeqn
This is a Hamiltonian system with the Hamiltonian functional
\be\label{hamilq0}
{\cal H}(\psi ,\pi ,q,p) =(1+p^2)^{1/2}+V(q)+\frac 12\int dx\Big(|\pi
(x)|^2+|\nabla \psi (x)|^2\Big) +\int dx\,\psi (x)\rho (x-q).
\ee
We have set the mechanical mass of the particle and the speed of wave
propagation equal to one.

We assume the real-valued function $\rho (x)$ to be smooth, radial symmetric, and
of compact support, i.e.
$$
\rho \in C^\infty_0({\R}^3)\,,\quad \rho(x)=\rho_1(r),\,\,\,r=|x|,\quad
\rho (x)=0\,\,\,\,\,\mbox{for}\,\,\,\,\,|x|\ge R_\rho >0 \,.\eqno{(C)}
$$

For the potential $V$ we require that it is sufficiently smooth and confining,
$$
V\in C^2(\R^3),\,\,\,
\lim\limits_{|q|\to\infty}=\infty.
\eqno{(P)}
$$

\subsection{Stationary solutions in the case of nonzero potential}

Let us define the set of stationary solutions to (\re{1}). For $q\in\R^3$ put
$$
\psi_q(x)=\int\fr{dy}{4\pi|y-x|}\rho(y-q).
$$
Let $S=\{q^*\in\R^3:\na V(q^*)=0\}$ be the set of critical points of $V$. Then the set ${\cal ST}$ of stationary solutions is given by
\be\la{stat}
{\cal ST}=\{(\psi,\pi,q,p)=Y_{q^*}:=(\psi_{q^*},0,q^*,0):q^*\in S\}.
\ee

\subsection{Soliton states in the case of zero potential}

Consider the nonperturbed system with $V\equiv 0$ corresponding to (\re{1}):

\beqn\la{system0}
\ba{ll}
\dot{\psi}(x,t)=\pi (x,t), & \dot{\pi}(x,t)=\Delta \psi (x,t)-\rho (x-q(t)),
\\
~ &  \\
\dot{q}(t)=p(t)/(1+p^2(t))^{1/2}, & \dot{p}(t)=\displaystyle\int dx\,\psi
(x,t)\,\nabla \rho (x-q(t))
\ea
\eeqn
with the Hamiltonian functional

\be\label{hamil0}
{\cal H}_0(\psi ,\pi ,q,p) =(1+p^2)^{1/2}+\frac 12\int dx\Big(|\pi
(x)|^2+|\nabla \psi (x)|^2\Big) +\int dx\,\psi (x)\rho (x-q).
\ee

The system (\ref{system0}) has the set of solutions that correspond
to the charge traveling
with uniform velocity, $v$. Up to translation
they are of the form
\be\la{tv}
S_v(t)= (\psi_v(x-vt),\pi_v(x-vt),vt,p_v)
\ee
with an arbitrary velocity $v\in V= \{v\in\R^3:\,|v|<1\}$.
The components of the travelling solution
can be calculated easily in Fourier transform, cf.  \ci{KS}:
\be\la{sol}
\psi_v(x) =
 -\fr  1{4\pi} \int
\fr {\rho(y)dy}
{|v(y-x)_\|+\lam(y-x)_\bot|}\,,\,\,\,\,
\pi_v(x) =-v\cdot\na\psi_v(x) , \,\,\, p_v= v/\lam.
\ee
Here we set
$\lam=\sqrt{1-v^2}$ and
$x= v x_\Vert+x_\bot$, where
$x_\Vert\in\R$
and $v\bot x_\bot\in\R^3$ for $x\in\R^3$; $p_v:=1/\sqrt{1-v^2}$. In further by ``solitons'' we mean
these travelling solutions to (\ref{system0}).

\begin{definition}\la{sm}
The soliton manifold for the system (\re{system0}) is the set
\be\la{sman}
{\cal SM}=\{(\psi_v(x-a),\pi_v(x-a),a,p_v):a\in\R^3,\,\,\,v\in V\}.
\ee
\end{definition}

In this note we study the property of global attraction, in the energy norm, of either the set ${\cal ST}$ of the stationary solutions to the system (\re{1}) or the soliton manifold ${\cal SM}$ for the system (\re{system0}). The precise definitions and theorems
to be stated in the following section.

\section{Existence of dynamics and main results}

First define a suitable phase space. Let $L^2$ be the real Hilbert space $L^2({\R}^3)$ with
norm $\br\cdot\br$, and let
 $\dot H^1$ be the completion of $C_0^\infty({\R}^3)$ with
norm
$\Vert\psi(x)\Vert=\br\nabla\psi(x)\br$.

\begin{definition}
The phase space ${\cal E}$ is the Hilbert space $\dot{H}^1\oplus
L^2\oplus {\R}^3\oplus {\R}^3$ of states
$Y=(\psi ,\pi ,q,p)$ with finite norm
\[
{\Vert \,Y\Vert }_{{\cal E}}=\Vert \psi \Vert +
\br\pi \br+|q|+|p|\,.
\]
\end{definition}

We write the
Cauchy problem for the system (\ref{1}) in the form
\begin{equation}  \label{evolut}
\dot{Y}(t)= {\bf F}(Y(t))\,,\quad t\in{\R}\,,\quad
Y(0)=Y^0\,,
\end{equation}
where $Y(t)=(\psi(t), \pi(t), q(t), p(t))$ and
$Y^0=(\psi^0, \pi^0,q^0, p^0)$.

\begin{pro}\la{ex}{\rm \cite{KSK,KS,KKS}}
Let  $(C)$ hold and either $(P)$ hold or $V\equiv0$,
 $Y^0= (\psi^0, \pi^0,q^0, p^0) \in {\cal E}$.
Then \\
{\rm (i)} The system (\re{1}) has
 a unique solution $Y(t)
=  (\psi(x,t),\pi(x,t), q(t), p(t)) \in C(\R,{\cal E})$
with $Y(0) =  Y^0$.\\
{\rm (ii)}
The energy is conserved, i.e.
 \be\la{2.4}
 {\cal H}(Y(t))= {\cal H}(Y^0)\,\,\,\,({\cal H}_0(Y(t))= {\cal H}_0(Y^0))\,\,\,for\,\,t\in\R.
 \ee
{\rm (iii)} The bound holds
\begin{equation}
\sup_{t\in {\R}}|\dot{q}(t)|\le \overline{v}<1\,,
\label{ov v}
\end{equation}
where $\overline{v}$ depends on $Y^0$ and
$\delta_\rho:=|\langle\rho,\Delta^{-1}\rho\rangle|
=\ds\int dk\,|\hat\rho(k)|^2|k|^{-2}$.
\end{pro}

The asymptotic behavior of solutions to (\re{1}) is by now well-known \cite{KSK,KS,KKS,IKS03,IKS12}. Either the set of the stationary solutions or the soliton manifold is the set of attraction of solutions, in some sense, under different conditions on $\rho$ and initial data. Anyway it was always mentioned \cite{KSK,KS,KKS,IKS03,IKS12} that by the energy conservation the convergence to a stationary state or a soliton cannot hold in the energy norm and could take place in local energy seminorms only. However, no example of the absence of convergence in the energy norm was presented.

In this note we study the question of global attraction, in the energy norm, of either the set of the stationary solutions or the soliton manifold. The question happens to be rather simple but, to our mind, fills a certain gap in the discussion.

\begin{definition}
i) The set of the stationary states ${\cal ST}$ has the property of global attraction in ${\cal E}$ if for any $Y_0\in{\cal E}$ $\lim\limits_{t\to\infty}{\rm dist}_{\cal E}(Y(t),{\cal ST})=0$, where $Y(t)$ is the solution to system (\ref{1}) with $Y(0)=Y^0$.

ii) In the case $V\equiv0$, the soliton manifold ${\cal SM}$ has the property of global attraction in ${\cal E}$ if for any $Y_0\in{\cal E}$ $\lim\limits_{t\to\infty}{\rm dist}_{\cal E}(Y(t),{\cal SM})=0$, where $Y(t)$ is the solution to system (\ref{1}) with $Y(0)=Y^0$.
\end{definition}

Our main (negative) result is the following theorem.

\begin{theorem}\la{attr}
i) For the system (\ref{1}), the set ${\cal ST}$ generally has no the property of global attraction in ${\cal E}$.

ii) For the system (\ref{system0}), in the case $V\equiv0$, the soliton manifold ${\cal SM}$ has no the property of global attraction in ${\cal E}$.
\end{theorem}

\section{Proof of Theorem \ref{attr}}

i) Let $V(q)=q^2$, this potential satisfies the condition (P). Then $q^*=0$. The corresponding $\psi_{q^*}=\psi_0$ reads in Fourier space as $\hat\psi(k) = -\hat\rho/k^2$. The unique stationary solution reads $S=(\psi_0,0,0,0)$. Its energy equals $H_0=\fr12\int\,dx\,|\na\psi_0|^2+\int\,dx\,\psi_0\rho$. For an initial condition $Y^0=(\psi^0,0,\pi^0,p^0)$ the energy equals $H^0=\sqrt{1+p_0^2}+\fr12\int\,dx(|\na\psi^0|^2+|\pi^0|^2)+\int\,dx\,\psi^0\rho$ and clearly we can choose a $Y^0$ such that $H^0>H_0$. Then by the energy conservation the solution with the initial condition $Y^0$ cannot converge to $S$ in ${\cal E}$.

ii) First let us compute the energy of the soliton $(\psi_v(x-a),\pi_v(x-a),a,p_v)$:
$$
H_{v,\,a}=\sqrt{1+p_v^2}+\frac 12\int dx\Big(|\pi_v(x)|^2+|\nabla \psi_v (x)|^2\Big)+\int dx\,\psi_v (x)\rho (x-a).
$$

We estimate the energy from below. Note that in Fourier representation one has
$$
\hat\psi_v=-\fr{\hat\rho}{k^2-(kv)^2},\,\,\,\hat\pi_v=-\fr{ikv\hat\rho}{k^2-(kv)^2},\,\,\,\sqrt{1+p_v^2}=\fr1{\sqrt{1-v^2}}\ge1.
$$
We have
$$
\int dx\,|\pi_v(x)|^2=\int dk\,|\hat\pi_v(k)|^2=\int dk\,\fr{-ikv\hat\rho\cdot(ikv\hat\rho)}{(k^2-(kv)^2)^2}=\int dk\,\fr{(kv)^2|\hat\rho|^2}{(k^2-(kv)^2)^2}\ge0.
$$
$$
\int dx\,|\na\psi_v(x)|^2=\int dk\,|-ik\hat\psi_v(k)|^2=\int dk\,\fr{-ik\hat\rho\cdot(-ik\hat\rho)}{(k^2-(kv)^2)^2}=\int dk\,\fr{k^2|\hat\rho|^2}{(k^2-(kv)^2)^2} \ge\int dk\,\fr{|\hat\rho|^2}{k^2}.
$$
$$
\int dx\,\psi_v(x)\rho(x-a)=\int dk\,\hat\psi_v(k)\overline{\widehat{\rho(x-a)}}=\int dk\,\hat\psi_v(k)e^{ika}\hat\rho(k)=
$$
$$
\int dk\,-\fr{\hat\rho}{k^2-(kv)^2}e^{ika}\hat\rho=-\int dk\,\fr{e^{ika}\hat|\rho|^2}{k^2-(kv)^2}.
$$
For the last term,
$$
|\int dx\,\psi_v(x)\rho(x-a)|\le\int dk\,\fr{|e^{ika}\hat\rho|}{|k|}\fr{|k|\hat\rho}{k^2-(kv)^2}\le\fr12\int dk\,\left(\fr{|\hat\rho|^2}{k^2}+\fr{k^2|\hat\rho|^2}{(k^2-(kv)^2)^2}\right).
$$
As the result,
$$
H_{v,\,a}\ge1-\fr12\int dk\,\fr{|\hat\rho|^2}{k^2}+\fr12\int dk\,\fr{k^2|\hat\rho|^2}{(k^2-(kv)^2)^2}.
$$
Since
$$
\int dk\,\fr{k^2|\hat\rho|^2}{(k^2-(kv)^2)^2}\ge\int dk\,\fr{|\hat\rho|^2}{k^2},
$$
we obtain the esimate
\be\la{ge1}
H_{v,\,a}\ge1
\ee

We try to construct initial data $(\psi^0,\pi^0,q^0,p^0)$ which energy $H^0$ is less than $H_{v,\,a}$, it would be enough that $H^0<1$. Put $\pi^0=0$, $q^0=0$, $p^0=0$, then $H^0 = 1+\frac 12\int dx|\nabla \psi^0(x)|^2+\int dx\,\psi^0(x)\rho(x)$ and it is sufficient to find a $\psi_0$ such that $\frac 12\int dx|\nabla \psi^0|^2+\int dx\,\psi^0\rho<0$.

Let us set $\psi_0=-\ve\rho$, then $\frac 12\int dx|\nabla \psi^0|^2+\int dx\,\psi^0\rho=\ve^2\frac 12\int dx|\nabla\rho|^2-\ve\int dx\,\rho^2<0$ for sufficiently small $\ve$.

Then the solution with the initial data $(-\ve\rho,0,0,0,)$ cannot be attracted by the soliton manifold in ${\cal E}$, by the energy conservation. \hfill $\Box$

\end{document}